\documentclass[a4paper]{spie}  

\usepackage[english]{babel}

\usepackage{amsfonts}
\usepackage{amsmath}
\usepackage{color}
\usepackage[colorlinks=true, allcolors=blue]{hyperref}

\usepackage{graphicx}

\newcommand{\onefig}[1]{\centering{\includegraphics[width=0.8\columnwidth]{#1}}}
\newcommand{\sclfig}[2]{\centering{\includegraphics[width=#2\columnwidth]{#1}}}

\usepackage{bm}

\newcommand{\xbas}{u}

\newcommand{\clvvec}{\gamma}

\newcommand{\euler}{\mathrm{e}}
\newcommand{\ramuno}{\mathrm{i}\mkern1mu}

\newcommand{\myprime}{^{\,\prime}}

\title{Estimating of the inertial manifold dimension for a chaotic
  attractor of complex Ginzburg-Landau equation using a neural
  network}

\author[a]{Pavel V. Kuptsov}%
\author[a]{Anna V. Kuptsova}%

\affil[a]{Institute of electronics and mechanical engineering, Yuri
  Gagarin State Technical University of Saratov, Politekhnicheskaya
  77, Saratov 410054, Russia}%

\authorinfo{Send correspondence to P.V.K. E-mail:
  p.kuptsov@rambler.ru}

\begin{document}

\maketitle

\begin{abstract}
  Dimension of an inertial manifold for a chaotic attractor of
  spatially distributed system is estimated using autoencoder neural
  network. The inertial manifold is a low dimensional manifold where
  the chaotic attractor is embedded. The autoencoder maps system state
  vectors onto themselves letting them pass through an inner state
  with a reduced dimension.  The training processes of the autoencoder
  is shown to depend dramatically on the reduced dimension: a learning
  curve saturates when the dimension is too small and decays if it is
  sufficient for a lossless information transfer. The smallest
  sufficient value is considered as a dimension of the inertial
  manifold, and the autoencoder implements a mapping onto the inertial
  manifold and back. The correctness of the computed dimension is
  confirmed by its remarkable coincidence with the one obtained as a
  number of covariant Lyapunov vectors with vanishing pairwise
  angles. These vectors are called physical modes. Unlike never
  having zero angles residual ones they are known to span a tangent
  subspace for the inertial manifold.
\end{abstract}


\section{Introduction}

Modeling and analysis of high dimensional nonlinear systems encounter
obvious obstacle related with large amount of data required to be
processed. However there are reasons to expect that much of these data
are actually spurious and can be eliminated. The simplest way to be
convinced of that is to compute Kaplan-Yorke dimension~\cite{KYDim}
that typically is much lower than the full phase space dimension. The
Kaplan-Yorke dimension is related with the information dimension and
is an upper estimate for the Hausdorff dimension of an
attractor~\cite{GrasProc83}. Hence it is natural to expect that actual
topological dimension of a manifold where the attractor is embedded is
at least not much higher. Rigorous arguments in favor of this can be
found in Ref.~\cite{InertManif} where the concept of inertial
manifolds is introduced. This manifold contains the attractor,
attracts exponentially all solutions, and is stable with respect to
perturbations. Even for infinite dimensional systems, like spatially
distributed ones, these manifolds are finite dimensional and the
dimension is not very high.

Obvious benefit of having the mappings to the inertial manifold and
back is that high or infinite dimensional models can be reduced to low
dimensional sets of equations without the loss of information. But
even if the low dimensional model is not constructed explicitly, the
lossless dimension reduction of state vectors eliminates spurious
information from data flow and can dramatically facilitate their
analysis.

General methods of finding the inertial manifold mappings are not
known yet, but there are approaches for building their
approximations~\cite{LiQi2010}. Moreover, there exist a lot of methods
for building approximate reduced models without explicit appealing to
the concept of inertial manifolds~\cite{Belkin2003,
  benner2005dimension, Kerschen2005, blum2013, Chorin9804}.

Power tools and methods for analysis of dynamical systems can be
borrowed and adopted from a rapidly growing area of machine
learning. For example, machine learning apparatus, particularly neural
networks, are employed for building computer models of spatially
distributed experimental systems~\cite{Zhou1996, Gonzalez1998, Li2004,
  Deng2005}. An example of such modeling includes dimension reduction
of experimental data via a nonlinear principal component analysis with
further creation a neural network model and spatiotemporal
reconstruction of the
solution~\cite{QiLi2009}. Papers~\cite{Wang2016A, Wang2016B} develop
an approach for building a low dimensional model of high dimensional
experimental data using an autoencoder neural
network~\cite{haykin2009neural}. First, dimension of the input data is
reduced using the encoding part of the autoencoder. Then based on the
reduced vectors a mathematical model is constructed. Finally, the
output of the functioning model is processed by the decoding part of
the autoencoder to predict original high dimensional data.

Series of papers~\cite{CLVInerManifPRL09, KupParStrictFussy,
  CLVInerManifPRE11, CLVInerManifPRL12} investigate inertial manifolds
of spatially distributed systems using covariant Lyapunov vectors
(CLVs). These vectors are one to one related to Lyapunov exponents and
point tangent directions of expanding, neutral and contracting
manifolds of attractor trajectories~\cite{GinCLV, WolfCLV,
  CLV2012}. Average exponential growth of perturbations along each CLV
is given by a corresponding Lyapunov exponent. As reported in
Ref.~\cite{CLVInerManifPRL09} for spatially distributed chaotic
systems there exists a sufficiently small subset of CLVs related to
all positive and several closest to zero negative Lyapunov exponents
that are highly entangled in the sense that angles between them often
vanish along trajectories. This subset of CLVs span a subspace tangent
to the inertial manifold and their number is equal to the dimension of
this manifold. These vectors carry all essential information about the
attractor and are called physical modes. In contrast the large
batch of residual CLVs are orthogonal to each other. These vectors are
tangent to trajectories approaching the attractor and called
spurious modes.

In this paper we determine the dimension of the inertial manifold for
a high dimensional chaotic system using an autoencoder. Then we
compute the dimension via CLVs as a number of physical modes. These
two values obtained using independent methods are found to be
remarkably identical. It confirms the correctness of these values and
open perspectives of using the developed approach based on machine
learning for further analysis of high dimensional chaotic dynamics.

The paper is organized as follows. In Sec.~\ref{sec:system} we
introduce complex Ginzburg-Landau equation whose attractor will be
considered. Section~\ref{sec:autoencoder} represents the
autoencoder. The dimension of the inertial manifold is determined here
via analysis of learning curves of this
network. Section~\ref{sec:autoecnprop} investigates properties of the
built autoencoder. It is shown that the autoencoder preserves the
distributions and Fourier spectra of the input signal and also is
robust with respect to small perturbation. In
Sec.~\ref{sec:dimviaclvs} we determine the dimension of the inertial
manifold using CLVs that coincide with the dimension obtained using
the autoencoder. Finally, in the concluding Sec.~\ref{sec:concl} we
summarize the results of the paper.

\section{Considered system}\label{sec:system}
\begin{figure}
  \onefig{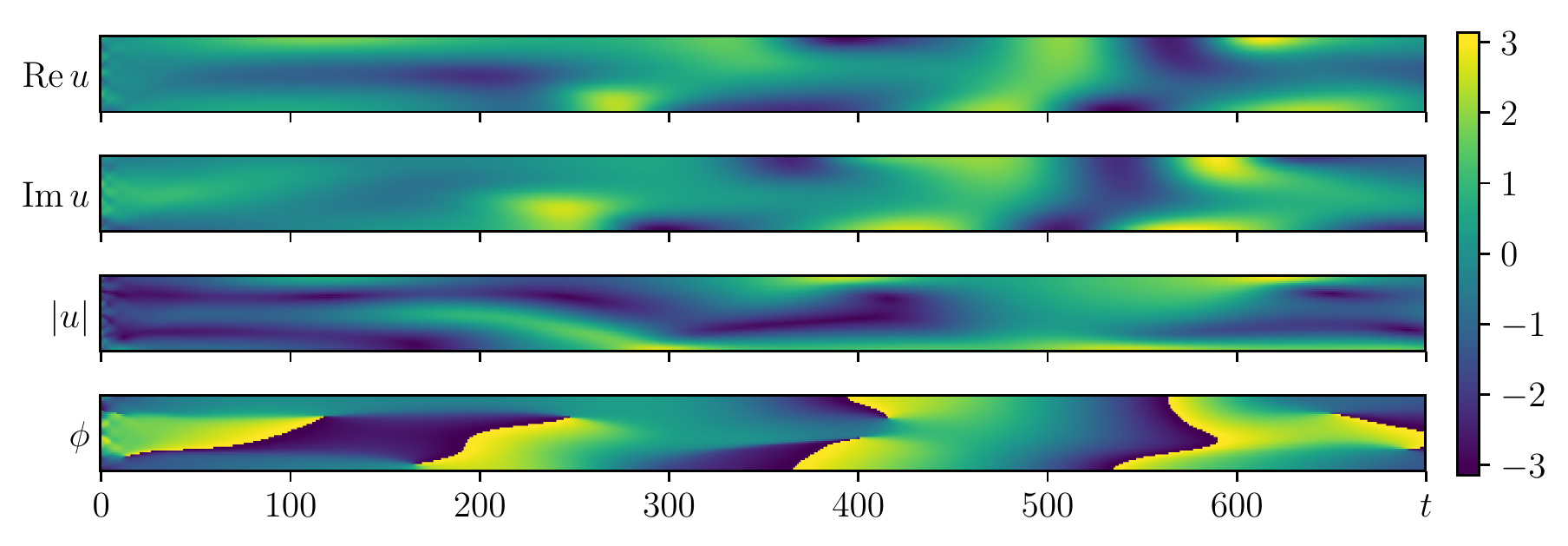}
  \caption{\label{fig:sptm}Spatio-temporal behavior of the
    system~\eqref{eq:cgle}. From bottom top to bottom: real part of $u$,
    imaginary part, amplitude and phase.}
\end{figure}

We are going to consider complex Ginzburg-Landau equation~\cite{CGLE}
\begin{equation}
  \label{eq:cgle}
  u_t = u - (1+\ramuno c_3) u|u|^2 + (1+\ramuno c_1) u_{xx}
\end{equation}
with the parameters values $c_3=3$, and $c_1=-2$ and boundary
conditions $u_x|_{x=0}=u_x|_{x=X}=0$. For numerical solution a mesh is
introduced with $N=40$ points and space step is $\Delta x=0.2$. It
corresponds to a spatial size $X=7.8$. The total dimension of the
discretized phase space is $D_u=80$. For these parameter settings the
system~\eqref{eq:cgle} demonstrates amplitude chaos as illustrates
Fig.~\ref{fig:sptm}.

The first six Lyapunov exponents of the system are $\lambda=0.18$,
$0.05$, $0.00$, $0.00$, $-0.23$, and $-1.17$. The corresponding
Kaplan-Yorke dimension~\cite{KYDim} is:
\begin{equation}
  \label{eq:dky}
  D_{\text{KY}}\approx 5.00.
\end{equation}
This dimension is known to be an upper estimate of the Hausdorff
(fractal) dimension of an attractor~\cite{GrasProc83} and we observe
that this is much smaller then the phase space dimension $D_u$. It
means that it is natural to expect that the attractor of the
system~\eqref{eq:cgle} lays on a manifold whose dimension is much less
then $D_u$.

\section{Autoencoder}\label{sec:autoencoder}

To find a manifold where the attractor of the system~\eqref{eq:cgle}
is embedded we will reduce dimension of its state vectors using an
autoencoder~\cite{haykin2009neural}. In brief, this is a neural
network that implements two successive transformations: encoding and
decoding. The encoder maps high dimensional data vectors to low
dimensional ones and the decoder maps these reduced vectors back to
the high dimensional space. The network is trained in such a way that
the initial and the reconstructed vectors coincide. As a result the
reduced vectors carry all information about the input high dimensional
data.

The encoding and decoding parts of an autoencoder can formally be
represented as vector functions $A$ and $B$, respectively:
\begin{equation}
  \label{eq:autoencoder_formulas}
  r = A(u), \; u\myprime = B(r).
\end{equation}
The encoding function $A$ maps vectors $u\in \mathbb{R}^{D_u}$ from
the discretized state space of the system~\eqref{eq:cgle} onto the
reduced space of vectors $r\in \mathbb{R}^{D_r}$, where $D_r$ is the
reduced dimension, $D_r<D_u$. The decoder $B$ maps the vectors $r$
back to the initial space of vectors $u\myprime \in \mathbb{R}^{D_u}$.
Training of the autoencoder~\eqref{eq:autoencoder_formulas}, i.e.,
tuning of its neuron weights, is performed in such a way as minimize
the loss function
\begin{equation}
  \label{eq:loss_funct}
  L=\|\vec u - \vec u\myprime\|^2/D_u.
\end{equation}
When $L\approx 0$ the decoder part $B$ approximates the inverse of the
function $A$, i.e., $B\approx A^{-1}$, so that the autoencoder as a
whole performs the identity mapping of the state vectors $u$ onto
themselves. The important point is that there is an intermediate
bottleneck where all the information is carried by the reduced vectors
$r$. Obviously such autoencoder can be trained for $D_r=D_u$, and when
$D_r$ is too small the training will not be successful. Thus we expect
that the minimal autoencoder~\eqref{eq:autoencoder_formulas},
\eqref{eq:loss_funct} with the smallest sufficient $D_r$ gives the
sought mapping onto the inertial manifold and back, and the dimension
of this manifold is $D_r$.

\begin{figure}
  \onefig{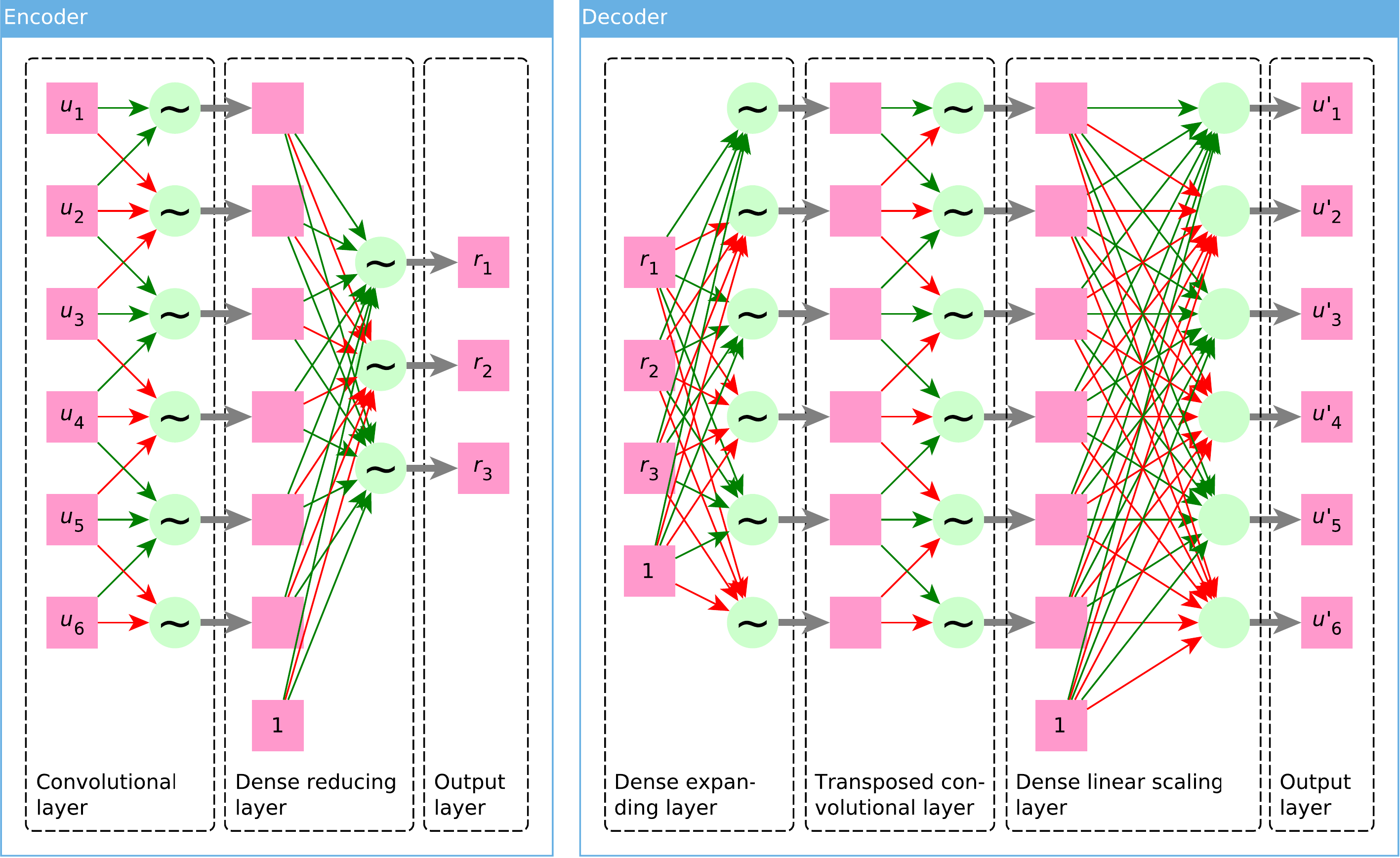}
  \caption{\label{fig:autoencoder}Graph representation of the
    autoencoder employed for dimension reduction of the
    system~\eqref{eq:cgle}. Arrows indicate interconnections of nodes.
    Circles with tilde symbol ``$\sim$'' inside represent an action of
    a nonlinear function $\tanh$ onto a summed input data. Empty
    circles (see dense linear scaling layer) stand for mere summation
    of the input data without a nonlinear transformation. Neuron
    weights are assumed to be assigned to each thin arrow. Squares
    with a unit symbol ``$1$'' inside are added to each dense layers
    to indicate constant biases added to a weighted neuron sums before
    further transformation. Fat horizontal arrows show data transfer
    between layers.}
\end{figure}

Particular architecture of the autoencoder can be different and must
be selected to achieve the best performance. In this paper we have
constructed the one represented in Fig.~\ref{fig:autoencoder}.  The
encoder includes convolutional and dense (fully connected) layers (see
Refs.~\cite{haykin2009neural,goodfellow2016deep,ConvArithm} for the
detailed explanation of these types of neural network
architectures). Using the convolutional layer is motivated by the
presence of the second spatial derivative in the model
equation~\eqref{eq:cgle} whose discretized version operates similarly
to the convolution. The convolutional layer accepts the initial state
vector $u$ and contains $D_u$ inputs and $D_u$ outputs. The reduction
occurs in the next dense layer with $D_u$ inputs and $D_r$
outputs. The outputs are elements of the reduced vector $r$.

The first layer of the decoder is dense. It accepts $D_r$ elements of
a vector $r$ and expand them to $D_u$ outputs that are further
processed by the transposed convolutional layer
(paper~\cite{ConvArithm} clearly explains the details of the
transposed convolution). The outputs of this layer is obtained as a
result of action of hyperbolic tangent and hence always belong to the
interval $(-1, 1)$. To recover the original scale of the data one more
dense layer is added that performs only linear scaling without a
subsequent nonlinear transformation. Its outputs form a vector
$u\myprime$.

The autoencoder is created using TensorFlow
framework~\cite{tensorflow2015-whitepaper} and the builtin
optimization algorithm Adam~\cite{Adam} is used for training. The
training set includes 30 trajectory cuts of the system~\eqref{eq:cgle}
each of 10000 vectors $u$. On each epoch of training the whole
training set is shuffled and split into 30 minibatches each of length
10000. These minibatches one by one are showed to the network and the
optimization step is performed. Quality of the training is estimated
on the validation set including 3 trajectories each of length
10000. For this purpose a learning curve is plotted showing how the
loss function $L$~\eqref{eq:loss_funct} computed for the validation
set behaves against the epoch number.

\begin{figure}
  \onefig{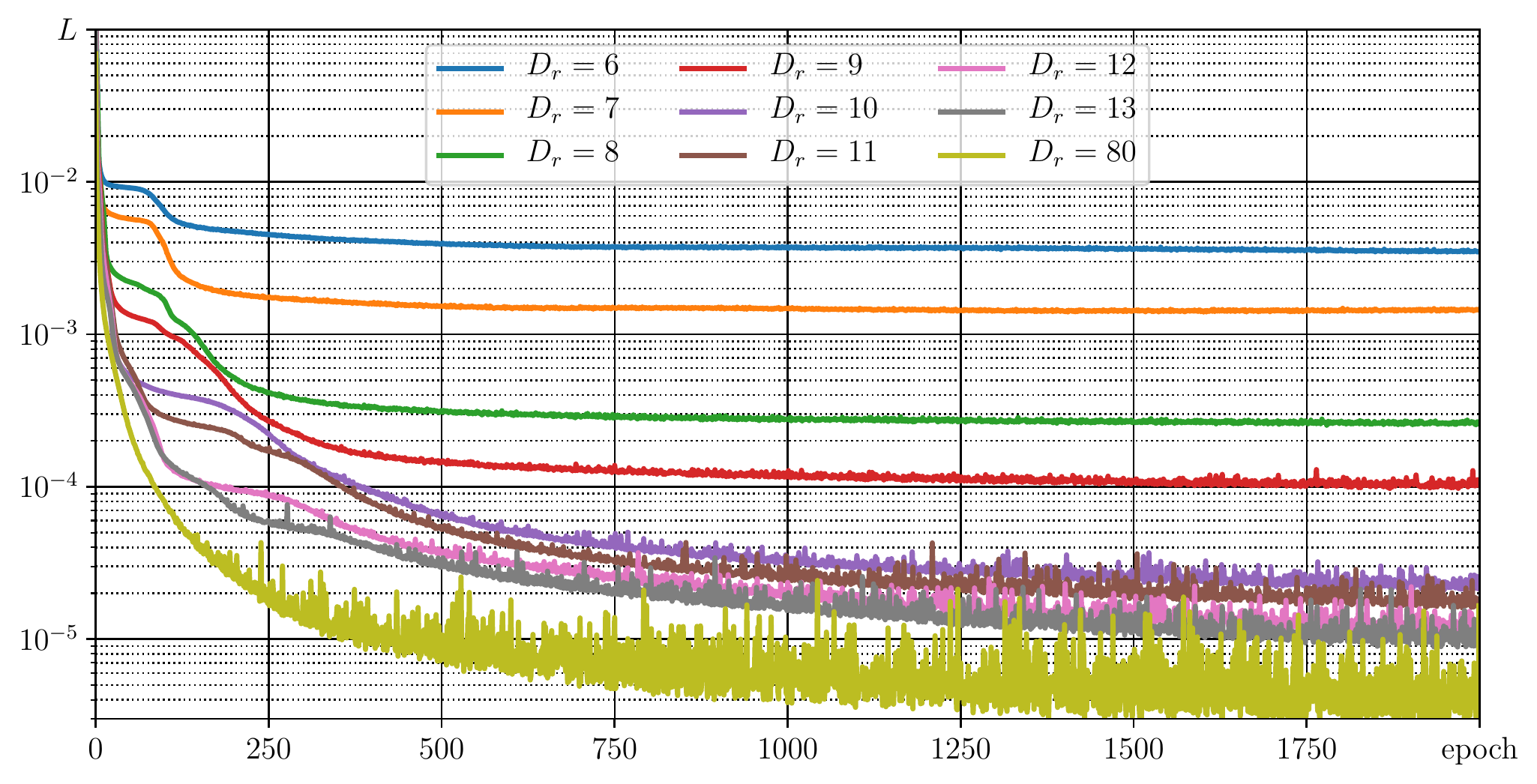}
  \caption{\label{fig:learncurves}Learning curves for different
    reduced dimensions $D_r$: the loss function~\eqref{eq:loss_funct}
    is plotted against epoch number. Semi-logarithmic scale is used.}
\end{figure}

Figure~\ref{fig:learncurves} demonstrates learning curves for
different $D_r$. One can see two types of them. The curves with
$D_r\leq 9$ reach saturation level and do not decays further, while
those with $D_r\geq 10$ continue decaying. On the basis of these
observations one can conclude that $D_r=10$ is the sought dimension of
the inertial manifold. This autoencoder is trained further up to 12000
epochs and the loss function level $10^{-5}$. Properties of the
mappings~\eqref{eq:autoencoder_formulas} implemented by the resulting
network are analyzed in the next section.

\section{Properties of the autoencoder}\label{sec:autoecnprop}

Inertial manifold in the original $D_u$-dimensional state space of the
system~\eqref{eq:cgle} is a $D_r$-dimensional smooth surface, probably
strongly curved. Due to the smoothness points located sufficiently
close to the attractor must remain close to it under the
mappings~\eqref{eq:autoencoder_formulas}. In the other words the
autoencoder have to be robust with respect to perturbations of input
data.

\begin{figure}
  \onefig{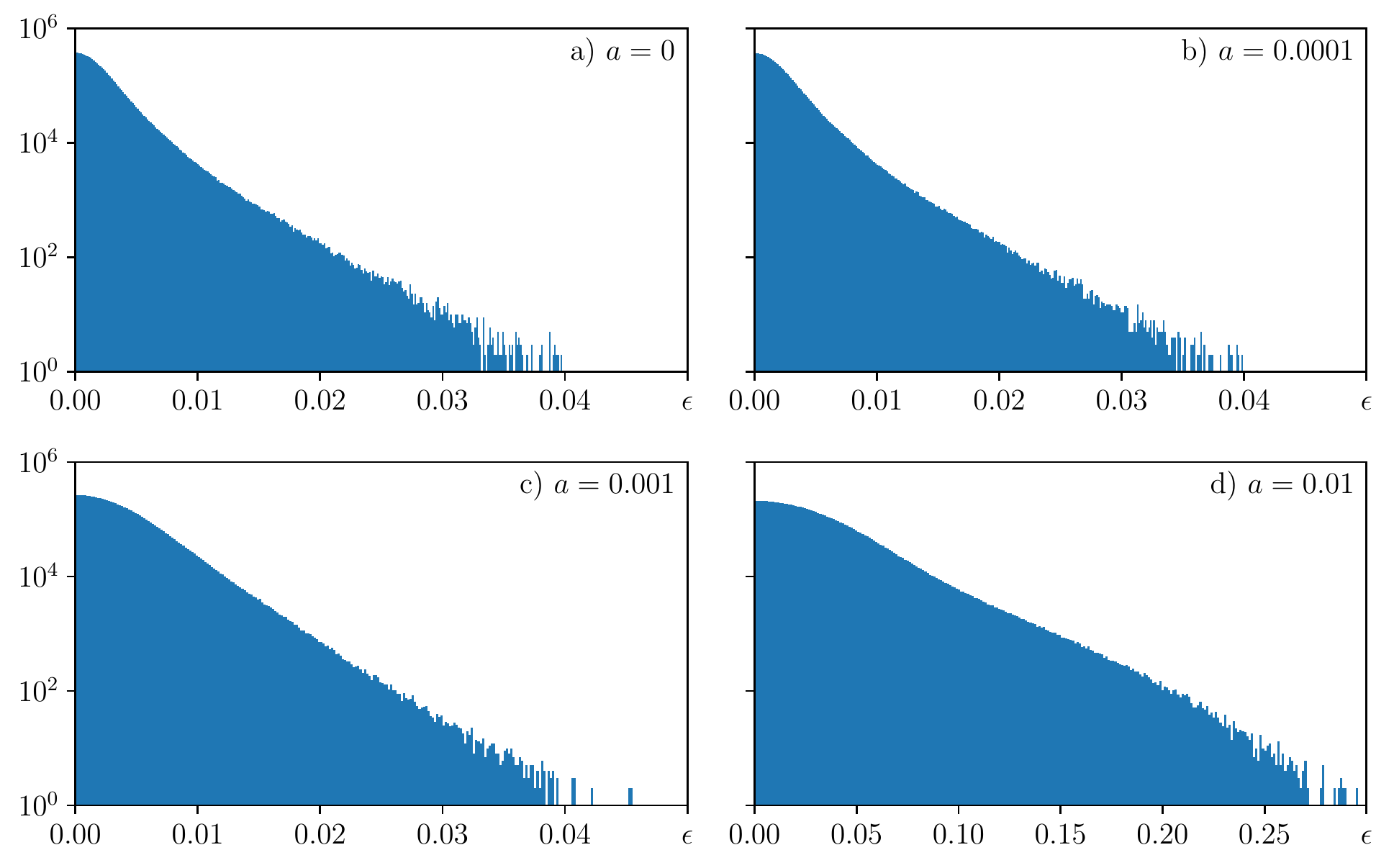}
  \caption{\label{fig:robustness}Reconstruction error distributions
    for perturbed attractor vectors $u$. Amplitudes of random
    perturbations $a$ are shown in the panels.}
\end{figure}

To verify the robustness, we collect totally $10^5$ attractor state
vectors $u$ from $10$ trajectories. Notice that these are new vectors
never seen by the network during the training. These vectors are
perturbed by adding to each site $u_i(t)$, $i=1,2,\ldots,D_u$, a
random variable with uniform distribution and amplitude $a$.This
perturbed batch is passed through the autoencoder and for each vector
site the absolute error of reconstruction is computed:
$\epsilon=|u_i(t)-u\myprime_i(t)|$. Distributions of this error are
shown in Fig.~\ref{fig:robustness}.

Figure~\ref{fig:robustness}(a) shows the distributions of
reconstruction error without adding a perturbation to the input
vectors, i.e., $a=0$, and in Figs.~\ref{fig:robustness}(b and c) the
perturbation amplitudes are $a=10^{-4}$ and $10^{-3}$,
respectively. One can see that in these three cases the error
distributions are almost identical and located below $0.05$. But
further increasing of the perturbation up to $a=0.01$, see
Fig.~\ref{fig:robustness}(d), results in dramatic change of the
distribution since the reconstruction error becomes much higher. These
observations demonstrates that the mapping onto the inertial manifold
and back remain valid when the perturbations remain small. It means
that the autoencoder is robust at least with respect to small
perturbations and recovers a smooth and strongly curved surface where
the attractor points are located.

\begin{figure}
  \onefig{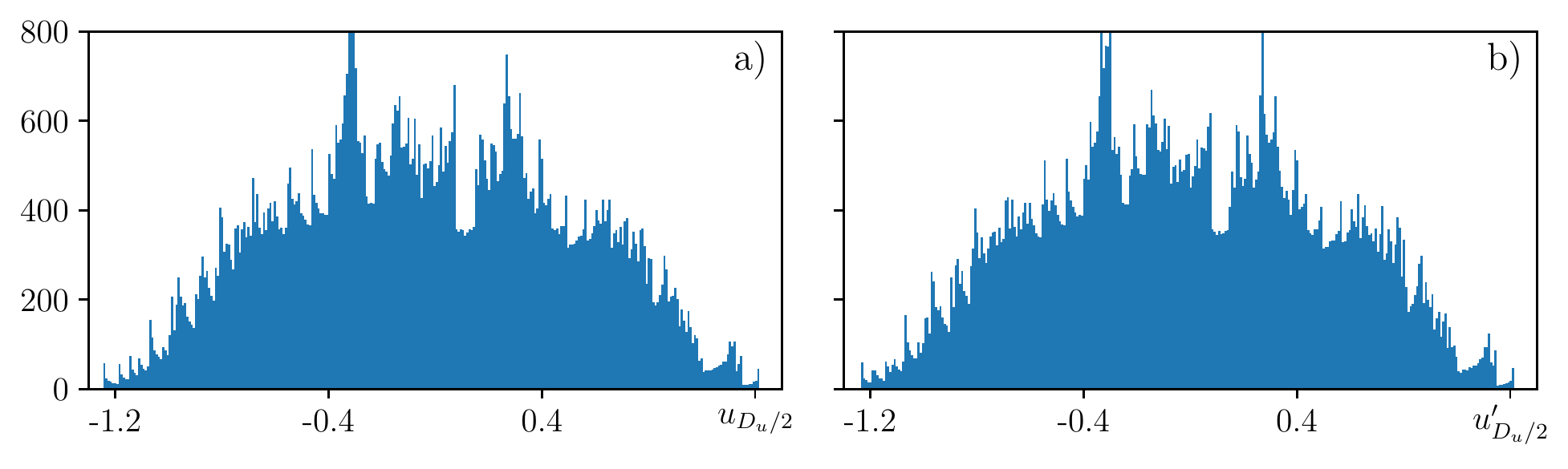}
  \caption{\label{fig:measure}Distributions of (a) $u_{D_u/2}(t)$ and
    (b) $u\myprime_{D_u/2}(t)$, i.e., elements of $u$ and $u\myprime$
    taken from their central sites $i=D_u/2$.}
\end{figure}

Visual image of operation results of the
mappings~\eqref{eq:autoencoder_formulas} can be obtained by
considering distributions on the attractor. First we again collect a
batch of $10^5$ attractor state vectors and let it pass through the
autoencoder. Then values at central site $u_{D_u/2}$ are gathered and
their distribution is plotted, see Fig.~\ref{fig:measure}(a). Also the
distribution of the reconstructed values $u\myprime_{D_u/2}$ is
plotted for comparison, see Fig.~\ref{fig:measure}(b). One can see
that these two distributions look identical.

It should be noted that each new run of the software computing data
for Fig.~\ref{fig:measure} results in a bit different distributions
since the sample length $10^5$ is insufficient to correctly represent
a high dimensional chaotic attractor of the
system~\eqref{eq:cgle}. However the important is that the initial and
reconstructed distributions are always identical.

\begin{figure}
  \onefig{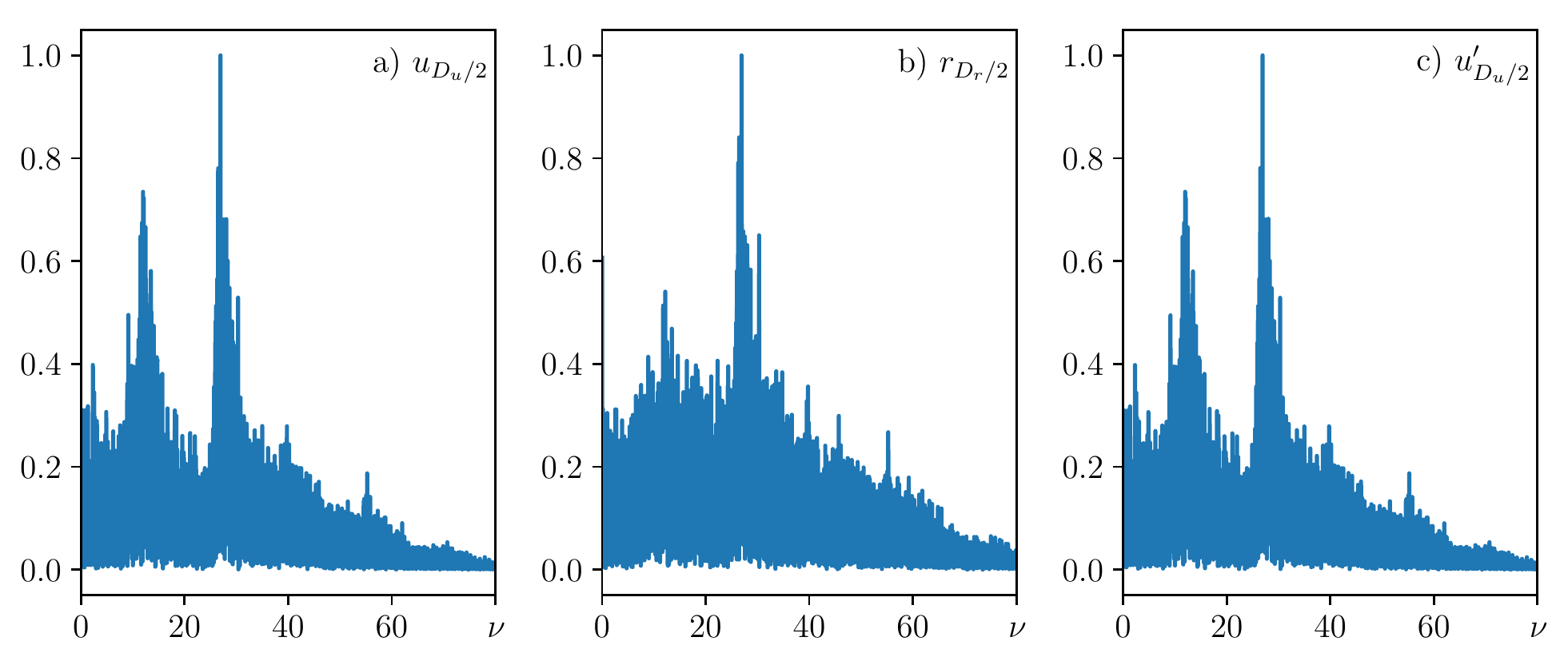}
  \caption{\label{fig:four}Fourier spectra of (a) $u_{D_u/2}(t)$, (b)
    $r_{D_r/2}(t)$, and (c) $u\myprime_{D_u/2}(t)$}
\end{figure}

One more test of the autoencoder is represented in
Fig.~\ref{fig:four}(a, b, and c) that show Fourier spectra for the
initial data $u_{D_u/2}(t)$, the reduced data $r_{D_r/2}(t)$, and the
reconstructed data $u\myprime_{D_u/2}(t)$, respectively. Observe very
high similarity of the spectra for the initial and reconstructed data:
in both panels there are two main harmonics surrounded by a chaotic
basis. Also notice that the reduced vector produces similar
spectrum. The highest main harmonics survives, and the second one
still exists though is much lower. Also similar are chaotic bases in
all three cases. It is important to note that the site number
$i=D_r/2$ where the data from the reduced vector are taken is chosen
arbitrary. The encoding transformation is performed so that no clear
correspondence between sites of the initial vector $u$ and the reduces
one $r$ can be established. In fact all sites of the reduced vector
demonstrates similar Fourier spectra.

\section{Dimension of the inertial manifold
  estimated via covariant Lyapunov vectors}\label{sec:dimviaclvs}

\begin{figure}
  \sclfig{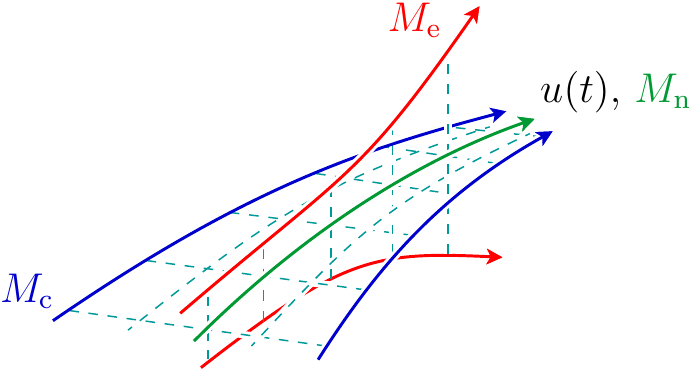}{0.4}
  \caption{\label{fig:manif}Attractor trajectory $\xbas(t)$ of a
    system and its manifolds: expanding $M_{\text{e}}$, neutral
    $M_{\text{n}}$, and contracting $M_{\text{c}}$. A continues time
    system always has at least one zero Lyapunov exponent that
    corresponds to translations along the trajectory. In this case the
    neutral manifold coincides with the trajectory.}
\end{figure}

Now we will compare the estimated dimension of the inertial manifold
$D_r=10$ with the one obtained via analysis of angles between
covariant Lyapunov vectors (CVLs). First of all we briefly recall what
are CLVs~\cite{GinCLV, WolfCLV, CLV2012}. Figure~\ref{fig:manif} show
a trajectory of a dynamical system $u(t)$ that belongs to a chaotic
attractor. Vicinity of such trajectory can be split into a direct sum
of expanding, neutral and contracting manifolds. All trajectories from
the contracting manifold approach the attractor; trajectories form the
expanding manifold approach it in the inverse time and the neutral
manifold is marginally stable. Average exponential expansion or
contraction ratios of these manifolds are given by Lyapunov exponents:
positive, zero and negative exponents correspond to expanding, neutral
and contracting manifolds, respectively. The numbers of these
exponents equal to the dimensions of the corresponding manifolds.

\begin{figure}
  \sclfig{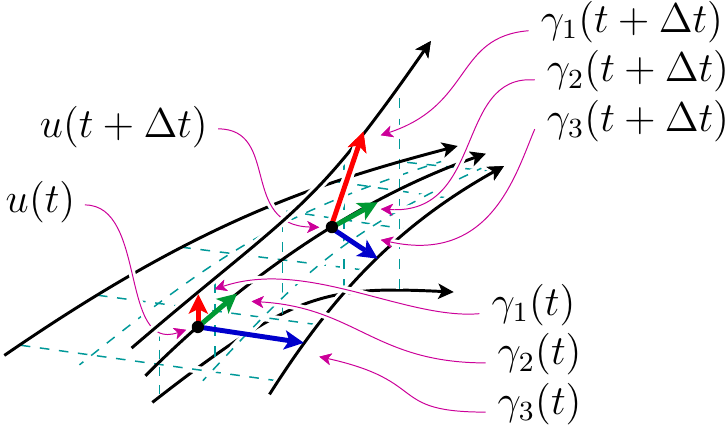}{0.4}
  \caption{\label{fig:manif_clv}CLVs $\clvvec_1$, $\clvvec_2$, and
    $\clvvec_3$ as tangent directions of trajectory manifolds.}
\end{figure}

CLVs are related one to one to Lyapunov exponents and show tangent
directions of the manifolds, see Fig.~\ref{fig:manif_clv}. One can see
here two trajectory points marked as $\xbas(t)$ and
$\xbas(t+\Delta t)$ and three CLVs, $\clvvec_1$, $\clvvec_2$, and
$\clvvec_3$, being tangent directions for expanding, neutral and
contracting manifolds. On average norms of CLVs grow or decay
exponentially as follows:
\begin{equation}
  \label{eq:clvs_in_time}
  \|\clvvec_i(t+\Delta t)\|=\euler^{\lambda_i\Delta t}\|\clvvec_i(t)\|,
\end{equation}
where $\lambda_i$ is $i$th Lyapunov exponent.

The important information about a dynamical system can be obtained
form the analysis of angles between its CLVs. The angles can be
computed in a straightforward way one by one. But also the efficient
algorithm for computations of angles between tangent subspaces spanned
by CLVs is suggested in Ref.~\cite{FastHyp12}.

Papers~\cite{CLVInerManifPRL09, KupParStrictFussy, CLVInerManifPRE11,
  CLVInerManifPRL12} investigate the possibility of recovering of the
inertial manifold by analyzing pairwise angles between CLVs. As
reported in Ref.~\cite{CLVInerManifPRL09} for highly dimensional
chaotic systems one can observe the splitting of the phase space into
two subspaces. The first one is spanned by the finite number of CLVs,
related with all positive, zero and several negative Lyapunov
exponents. These vectors, called physical modes, are highly entangled
in the sense that their angles often vanish. The second tangent
subspace is spanned by the CLVs, called spurious modes, related with
the residual negative Lyapunov exponents. Unlike the first ones, these
CLVs never have zero angles with each other. As argued in
Ref.~\cite{CLVInerManifPRL09}, physical modes are responsible for the
essential properties of the attractor. In particular, their number
equals to the dimension of the inertial manifold. The spurious ones
are only responsible for directions approaching the attractor. As
noted in Ref.~\cite{KupParStrictFussy}, additionally in the very end
of the CLVs spectrum there is one more batch of entangled vectors
related with the most negative CLVs. These vectors appear due to time
reversibility and become physical modes when the system moves backward
time.

\begin{figure}
  \sclfig{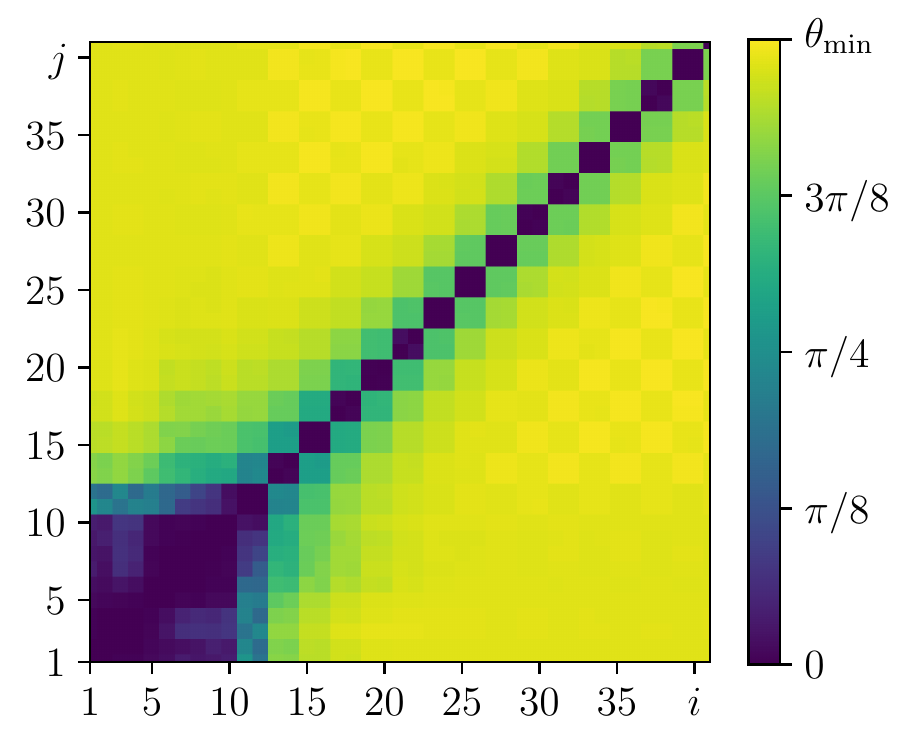}{0.4}
  \caption{\label{fig:pairang}Pairwise minimal angles between first
    $40$ CLVs computed after running along $10^6$ attractor points.}
\end{figure}

Results of searching of the physical modes and the estimation of the
inertial manifold for the system~\eqref{eq:cgle} via CLVs is shown in
Fig.~\ref{fig:pairang}. One can see here minimal pairwise angles
between first $40$ CLVs. First of all notice that the first $10$
vectors are strongly entangled, i.e., their angles can vanish. It is
this vector collection that form a set of physical modes. Their number
$10$ agrees with our previous estimation $D_r=10$ of the dimension of
the inertial manifold obtained via training of the autoencoder, see
Fig.~\ref{fig:learncurves}. The residual CLVs form spurious
modes. Dark $2\times 2$ squares on the main diagonal indicate that
they are entangled by pairs. This is related with the fact the
system~\eqref{eq:cgle} is complex valued so that neighboring CLVs are
related with real and imaginary parts of the dynamical variable.

\section{Conclusion}\label{sec:concl}

In this paper we build a neural network to find the smallest dimension
sufficient for lossless representation of the attractor of high
dimensional chaotic system. This representation is known as inertial
manifold. The network we build is an autoencoder whose encoding part
maps attractor vectors from the original phase space onto the inertial
manifold with a reduced dimension, and the decoding part performs the
inverse mapping back to the original space. As an example complex
Ginzburg-Landau equation is considered in a chaotic regime. Its phase
space after numerical discretization has dimension $80$, while the
reduced dimension is found to be $10$. The reduced dimension is found
after probing its different values and analyzing behavior of
corresponding learning curves. If the dimension is too small to
provide the lossless transformation, the learning curve tends to
saturation. Thus the appropriate reduced dimension is chosen as the
smallest one for that the learning curve still decays. The built
autoencoder is robust with respect to small perturbations. It means
that small perturbations to the attractor vectors remain small after
passing through it. Moreover the autoencoder preserves Fourier spectra
of the attractor trajectories and distributions of vector elements.

The computed dimension of the inertial manifold is compared with the
one obtained as a number of covariant Lyapunov vectors with vanishing
angles between each other. These vectors are called physical modes and
span a subspace tangent to the inertial manifold. The dimensions in
both cases perfectly coincide. It confirms the correctness of the
performed analysis and open perspectives of further development of
machine learning methods for studying high dimensional chaotic
systems.

\bibliography{aedim}
\bibliographystyle{spiebib} 

\end{document}